\documentclass[aps,preprintnumbers,amsmath,amssymb,twocolumn]{revtex4}

\usepackage{graphicx}
\usepackage{dcolumn}
\usepackage{bm}
\usepackage{color}
\usepackage{amssymb}   

\begin{document}

\title{Shear-induced migration in colloidal suspensions}

\author{Philip Marmet}
\author{Alberto Scacchi}
\author{Joseph M. Brader}\email {joseph.brader@unifr.ch}

\affiliation{
Department of Physics, University of Fribourg, CH-1700 Fribourg, Switzerland
}

\date{\today}

\begin{abstract}
Using Brownian dynamics simulations we perform a systematic investigation of the 
shear-induced migration of 
colloidal particles subject to Poiseuille flow in both cylindrical and planar geometry. 
We find that adding an attractive component to the interparticle interaction 
enhances the migration effect, consistent with recent simulation studies of platelet 
suspensions.
Monodisperse, bidisperse and polydisperse systems are studied over a range of shear rates, 
considering both steady states and the transient dynamics arising from the onset of flow. 
For bidisperse and polydisperse systems size segregation is observed.
\end{abstract}
\maketitle


\section{Introduction}

Soft materials, such as colloidal suspensions, display a wide range of response to externally 
applied deformation. Even in the case of simple shear flow one can observe nonlinear rheological 
phenomena such as shear thinning, shear thickening and yielding \cite{brader_review}, as well as 
shear-induced spatial inhomogeneities, such as shear banding 
\cite{banding1,banding2,banding3,banding4} or 
lane formation \cite{brader_kruger,lane_brady}. 
Simple shear is characterised by a translationally invariant shear-rate, reflecting 
the fact that only the relative velocity of a colloid with respect to its surroundings is of physical 
relevance, but can nevertheless generate interesting flow instabilities, such as those 
arising from shear gradient-concentration coupling in dense systems 
\cite{banding2,banding3,banding4}.
Situations for which the shear-rate is a function of position are in general less well 
understood, because the spatially dependent shear-rate couples in a complicated 
way to both the density and stress in the nonequilibrium system. 

In addition to the aforementioned phenomena, an important new physical mechanism is 
present in systems with a spatially varying shear-rate: shear-induced migration. 
If the shear-rate varies significantly on the scale of 
a colloid diameter, then the particles undergo a biased diffusion which causes them to drift 
(`migrate') to regions of lower shear gradient. 
The migration of colloids in channel flow is a dominant transport 
mechanism in suspensions, which has been exploited to facilitate segregation 
in colloidal mixtures \cite{mig1,mig2} and which is relevant for understanding the flow of 
blood \cite{mig2} as well as for numerous applications, such as food processing \cite{mig3}. 
Moreover, microfluidic devices in which small quantities of fluid are driven through a 
microchannel are an emerging technology for processing suspensions (e.g. pharmaceuticals). 

The present understanding of migration physics has its origins in the 1980 study of 
Gadala-Maria and Acrivos~\cite{gadala}, who performed viscosity measurements on a suspension of 
large (non-colloidal) spheres using a Couette rheometer. For volume fractions 
above approximately $30$\% they observed 
an anomalous drift in the measured viscosity as a function of time, which eventually 
saturated to a reproducible steady-state value. 
It was only several 
years later that Leighton and Acrivos proposed a microscopic mechanism for 
the observed time-dependence of the viscosity~\cite{mig4}. 
They realized that for flows with a spatially dependent shear-rate the collision frequency 
of a particle with its neighbours is not isotropically distributed over the surface of the particle - 
the surface regions subject to a higher shear-rate will experience on average a greater number of 
collisions than those subject to a lower shear-rate. This leads on sufficiently large time-scales 
to a biased diffusive motion. 
For the Couette rheometer employed by Gadala-Maria and Acrivos~\cite{gadala} a shear 
gradient can be found in the region between the bottom of the rotating cylinder and the (static) 
fluid reservoir underneath it. This drives a steady migration of particles from the gap between the 
two Couette cylinders into the reservoir, a process which continues until the migration flux is balanced by 
the flux due to the particle concentration gradient.

Since the original phenomenological theory of Leighton and Acrivos~\cite{mig4} there have been a 
variety of theoretical, simulation and experimental studies addressing migration physics 
\cite{nozieres,mig5,scacchi_krueger_brader2016,mig6,mig7,mig8,mig9,mig10}. 

The earliest of these studies were the phenomenological theories of Nozieres {\it et al.}~\cite{nozieres} and Phillips {\it et al.}~\cite{mig5}. The work of phillips extended the methods of Leighton and Acrivos to enable a self consistent 
determination of the density and velocity fields in a flowing suspension. 
More recently, microscopic density functional approaches have 
provided a route to calculate directly from 
the interparticle interaction potential the flow-induced changes in the 
density~\cite{brader_kruger,scacchi_krueger_brader2016}. 
The migration of non-Brownian particles under pressure-driven flow has been simulated, 
using Stokesian dynamics, 
by Nott and Brady~\cite{mig6} and Morris and Brady~\cite{mig7}. 
On the experimental side, nonequilibrium density distributions have been measured 
using confocal microscopy by Frank {\it et al}~\cite{mig9} and laser-Doppler velocimetry by 
Lyon and Leal~\cite{mig10}. 
All of the above-mentioned works focus on monodisperse systems. 
Bidisperse systems have been studied experimentally by Lyon and Leal~\cite{mig11} and by 
Semwogerere and Weeks~\cite{mig1,mig2}, both of whom observed segregation between the two 
species.


In this paper we present a systematic Brownian dynamics simulation study of particle 
migration in systems under Poiseuille flow. 
We begin by focusing on the simplest case: monodisperse, repulsive spheres. 
The dependence of both the steady state density and the transient dynamics on the shear-rate 
are investigated and we explore the differences between cylindrical (pipe) and planar (channel) 
geometries, an aspect which has not yet been addressed. 
When considering both bidisperse and polydisperse systems we identify 
segregation effects, which could potentially be exploited to separate particles according 
to their size. 
Building upon these results for repulsive systems we proceed to investigate the migration 
of attractive Lennard-Jones particles, both monodisperse and polydisperse. 
The addition of an attractive component to the pair potential is found to significantly enhance 
the strength of the migration effect. With the exception of the study of Katanov {\it et al.} 
\cite{blood}, in which a detailed model of blood flow was simulated, we believe that this work 
represents the first study of migration in systems with an attractive interaction potential.

The paper is organized as follows: In section \ref{numerical_algorithm} we give details of 
our Brownian dynamics simulation algorithm and in section \ref{results} we present the 
numerical results thus obtained. 
In \ref{mono_disperse} we consider monodisperse repulsive spheres subject to Poiseuille flow 
in both cylindrical and planar (channel) geometries. Both steady states and transient dynamics are 
addressed. 
In section \ref{poly_disperse} we generalize to bidisperse and polydisperse 
repulsive spheres and identify segregation effects. 
In \ref{mono_disperse_LJ} we consider the influence on migration of an attractive 
component to the pair potential, addressing the transient dynamics and steady-states 
of Lennard-Jones particles.  
In \ref{poly_disperse_LJ} we generalize to polydisperse Lennard-Jones spheres.   
Finally, in section \ref{conclusion}, we discuss our findings and give an outlook. 

\begin{figure}[!t]
\centering			
\includegraphics[width=0.51\textwidth]{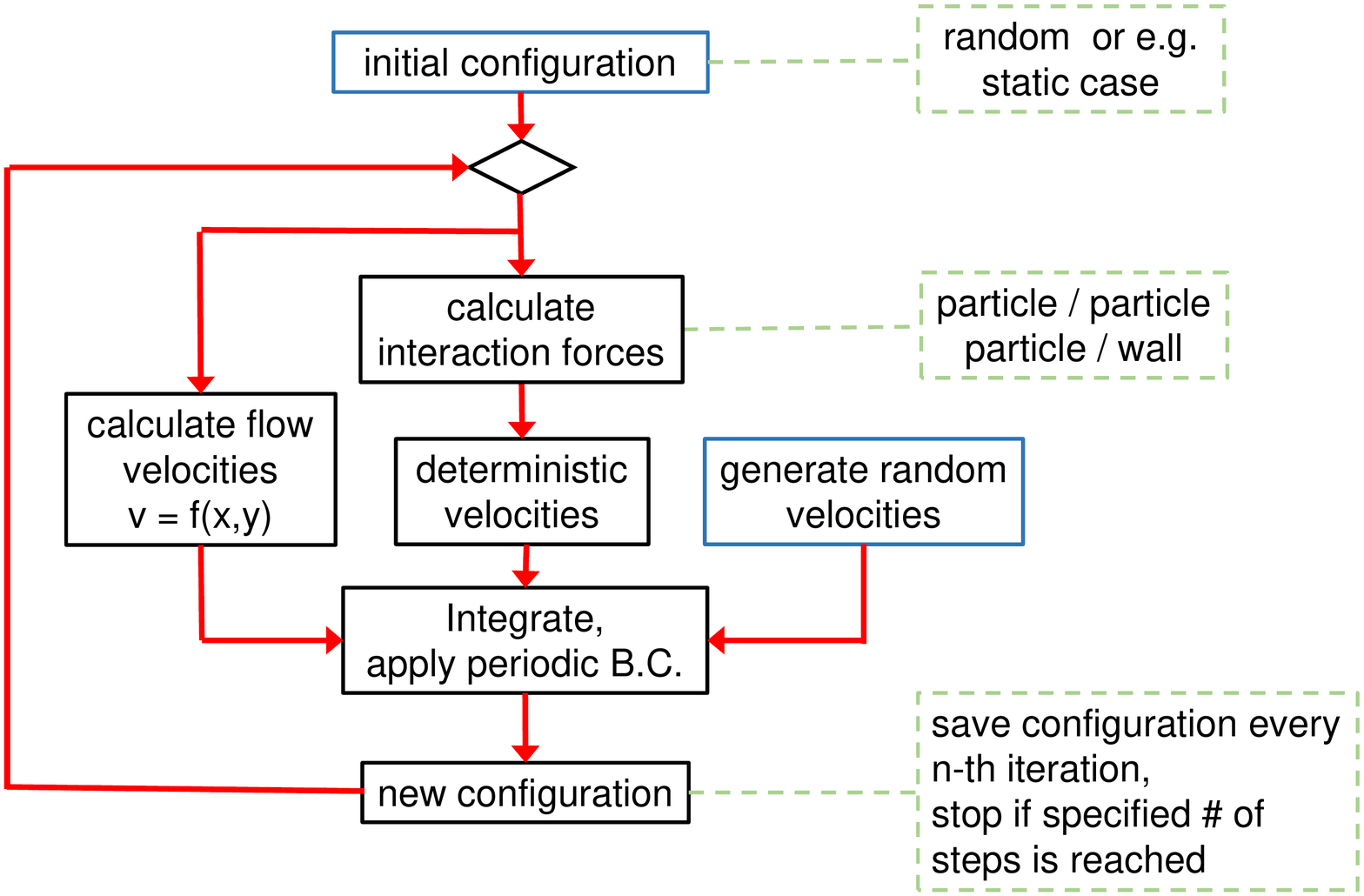}
\caption{Schematic representation of the Brownian dynamics algorithm.}
\label{BD_scheme}
\end{figure}

\section{Numerical algorithm}\label{numerical_algorithm}

We perform Brownian dynamics simulations of $N$ particles, randomly initialized without 
overlap \cite{allen}. 
The random velocity used to generate the thermal Brownian motion is sampled from a Gaussian 
distribution with standard deviation 
$\sigma=\sqrt{2 T'dt}$, where $dt$ is the Brownian timestep. The parameter $T'$ is given 
by $T'=k_B T/3\pi\eta d$, where $k_B T$ is the thermal energy, $\eta$ is the fluid 
viscosity and $d$ is the particle diameter.
For the pipe geometry the system is confined to a cylinder with periodic boundary conditions 
along the cylinder axis, whereas for the slit geometry we use periodic boundaries in the 
two directions parallel to the wall.  

 
In order to model the flow of a suspension driven either down a cylindrical pipe or through a slit, 
we add to the random Brownian velocities a deterministic contribution taken from the 
Poiseuille velocity profile appropriate to the geometry of interest. For the cylindrical 
geometry, for example, we employ
\begin{align}
v(r)=v_{\rm m}\left(
1 - \left(\frac{r}{R}\right)^2
\right),
\end{align}
where $v_m$ is the maximal velocity at the centre of the pipe and $R$ is the 
pipe radius. We assume that shear gradients are small on the length scale of the colloids.
While using a pre-determined quadratic velocity profile to model pressure-driven flow 
is convenient, for finite colloidal volume fractions it represents an approximation: the 
nontrivial coupling between the stress, strain-rate and density fields will in reality 
lead to a `blunting' of the velocity profile and thus a deviation from Poiseuille. 
However, provided that the colloidal volume fraction remains sufficiently low, then the 
deviation of the true velocity profile from the Poiseuille form will be minimal. 

Fortunately for the present work, the shear-induced migration effect can be well studied 
within the low volume-fraction regime. 
We thus work with a colloidal volume fraction of $\phi=0.05$ and, for convenience, 
we take the system temperature $T=1$. For simulations performed in cylindrical geometry  
we will fix the pipe radius as $R=10\,d$. 
All times will be given in units of the diffusion time of a single particle, namely 
in units of $\tau_{\rm B}=d^2/D_0$, where $D_0$ is the bare diffusion coefficient. 
The steps of the numerical algorithm for our Brownian dynamics simulations 
are summarised schematically in Fig.~(\ref{BD_scheme}).

\begin{figure}[!b]
\centering			
\includegraphics[width=0.5\textwidth]{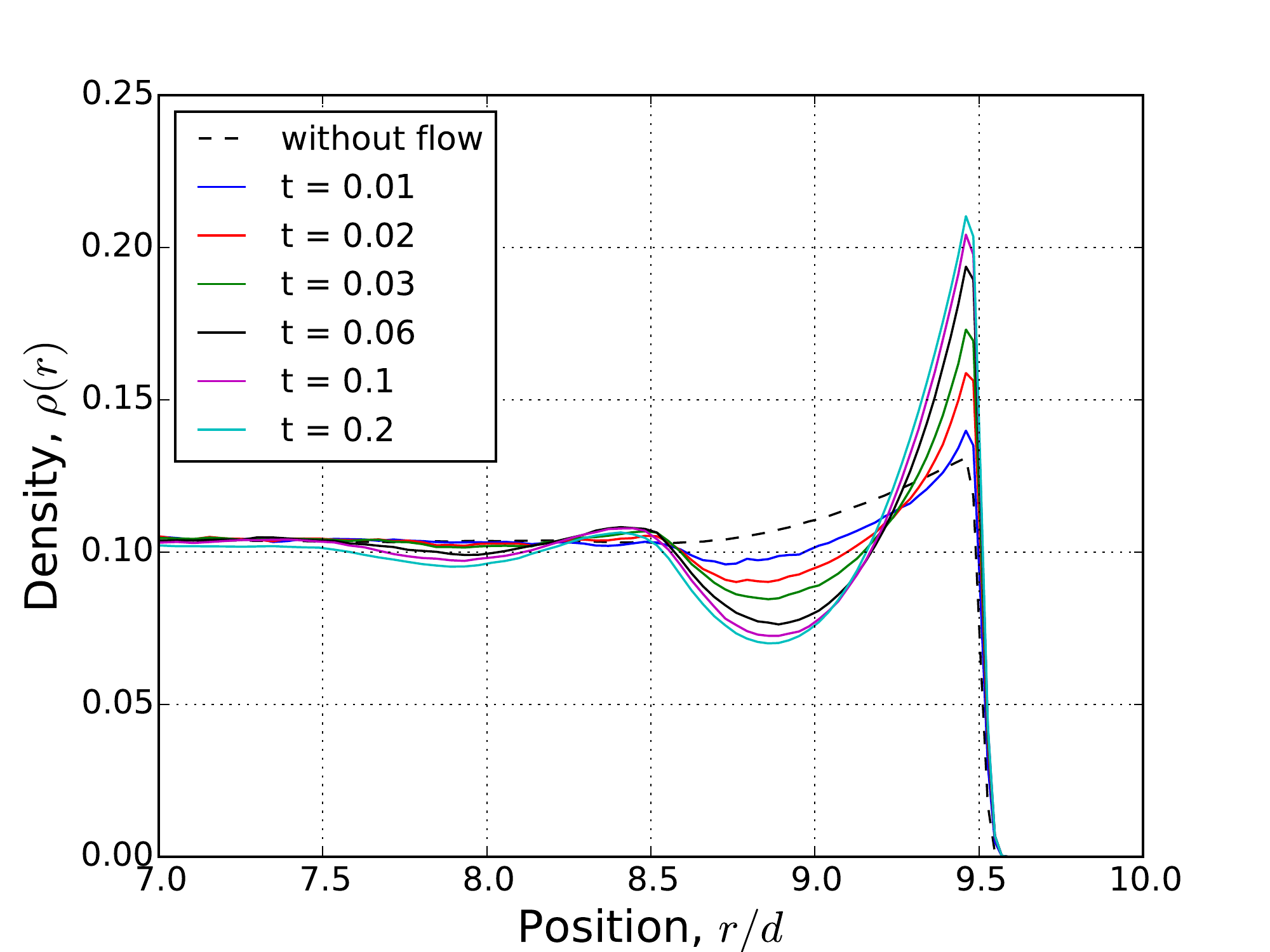}
\caption{Density profiles at the wall for monodisperse repulsive spheres 
confined by a cylindrical repulsive substrate (pipe geometry). We show the density profile for different 
times following the onset of Poiseuille flow at Peclet number $Pe=250$.
}
\label{rho_pipe_phi_0_05_Pe1250_Mono_hSph_f_t_sec1}
\end{figure}

\section{Results}\label{results}
\subsection{Monodisperse repulsive spheres}\label{mono_disperse}
We study first the nonequilibrium density profile of a system of $N$-colloids confined 
in a cylindrical geometry. 
For simplicity hydrodynamic interactions are neglected.
The interparticle interaction we choose to model our soft spheres 
is given by
\begin{equation}
U_{HS}=\begin{cases}4\epsilon\left[\left(\frac{d}{r}\right)^{24}-\left(\frac{d}{r}\right)^{12}+\frac{1}{4}\right]&$,   $r\leq 2^{\frac{1}{12}}d \\
0&$,   $r > 2^{\frac{1}{12}}d\end{cases}\label{UHS_potential}
\end{equation}
where the parameter $\epsilon$ has been set to $\epsilon=1$. The interaction between the 
colloidal suspension and the confining walls is given by the potential form 
(\ref{UHS_potential}), with the radial coordinate, $r$, replaced by 
the distance between the center of the particle and 
the substrate. The simulation variables have been set as follows: the 
number of particles $N=400$, the Brownian time step $dt=2\times 10^{-5}$ and the length of the 
pipe $L=13.33d$.
\\
We first focus on the time-evolution of the average density profile from equilibrium to 
steady-state, following the onset of flow. 
In order to provide a dimensionless characterization of the strength of the flow, we 
define the Peclet number as $Pe=(v_{m}d^2)/(R D_{0})$. 
In Fig.~(\ref{rho_pipe_phi_0_05_Pe1250_Mono_hSph_f_t_sec1}) we show the development of the density 
distibution close to the pipe wall as a function of density. 
As time progresses, the height of the first peak (closest to the wall) increases steadily before 
saturating to the steady-state value. This reflects an increase in the pressure acting on the wall 
and is a consequence of interparticle collisions pushing particles up against the confining boundary. 
Closely assocciated with the growth of the first peak is the development of an enhanced oscillatory 
packing structure - the flowing system has a tendency to form layers (concentric rings 
in the present geometry) of particles so as to minimize collisions. 
We note that a nonequilibrium packing structure close to the wall develops rather quickly following 
the onset of flow, with an approximate local steady-state being established by $t\approx 0.2$ 
(some smaller changes are observed to occur at later times, see below). 

\begin{figure}[!t]
\centering			
\includegraphics[width=0.5\textwidth]{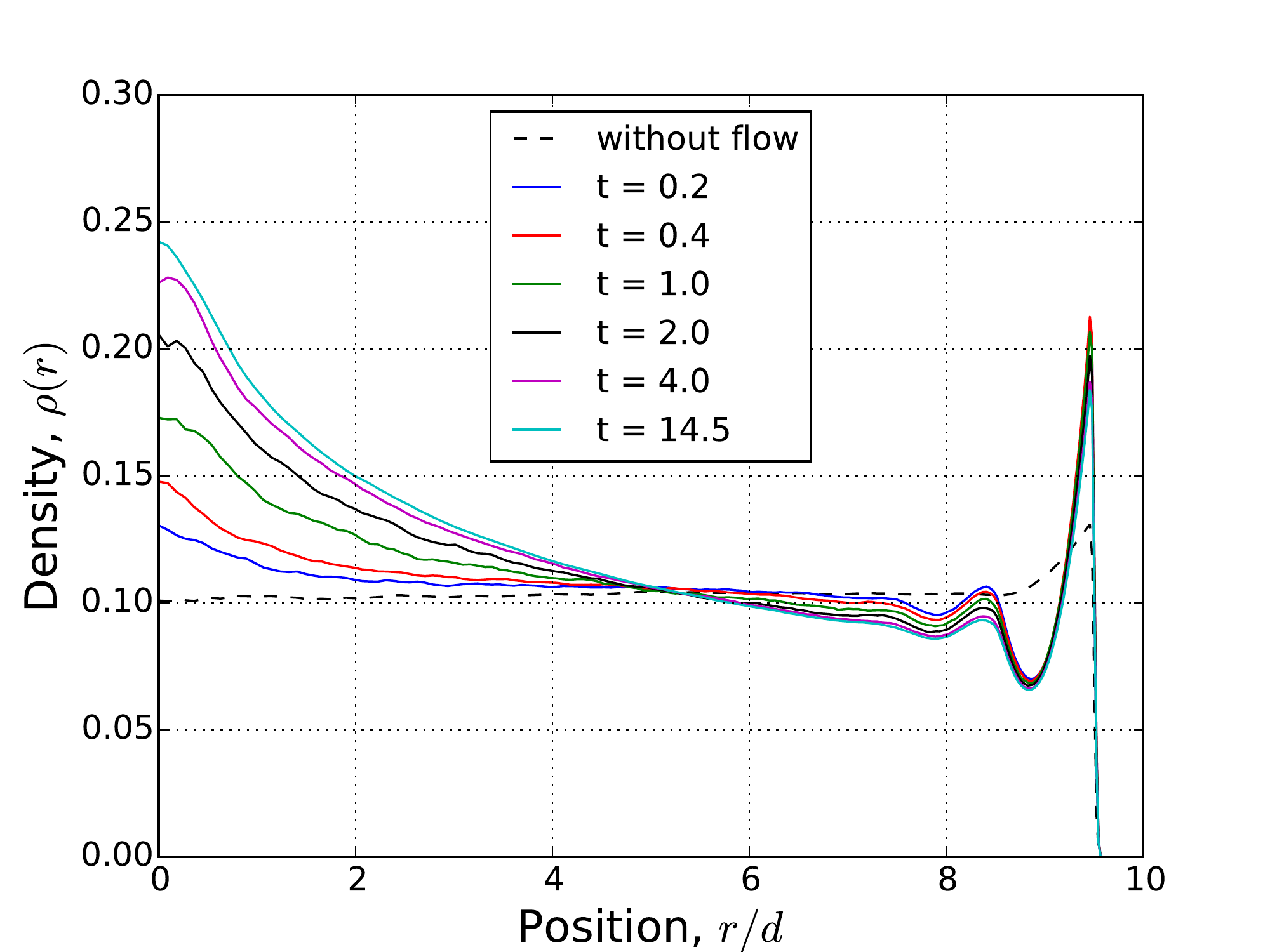}
\caption{As in Fig.~(\ref{rho_pipe_phi_0_05_Pe1250_Mono_hSph_f_t_sec1}), but focusing on the central 
peak region (around $r=0$) and for later times following the onset of flow.
} 
\label{rho_pipe_phi_0_05_Pe1250_Mono_hSph_f_t_sec2}
\end{figure}

In contrast to the rapid density changes observed close to the wall, the migration of particles 
in the non-uniform Poiseuille flow occurs much more slowly. In Fig.~(\ref{rho_pipe_phi_0_05_Pe1250_Mono_hSph_f_t_sec2}) 
we show the time-evolution of the full radial density profile. 
The earliest density profile shown in the figure is for $t\!=\!0.2$, by which time the structure 
close to the wall is approximately in steady state. As time progresses further we observe a gradual 
drift of particles towards the centre of the pipe, driven by the migration mechanism described earlier. 
The migration peak at the centre of the density profile eventually saturates at around $t\!=\!15$. 
The growth of this peak is a slow process, which requires material to be transported from the 
periphery of the pipe towards the centre. 
This transport depletes the density in the wall region and leads to slight modification of the 
oscillatory structure of the profile with increasing time. 
As the migration peak builds up, the emerging density gradient starts to counteract the migration 
force until a steady-state balance is eventually achieved. 
We find that the growth of the central peak as a function of time can be well approximated by 
an exponential. 


%

\begin{figure}[!b]
\centering			
\includegraphics[width=0.5\textwidth]{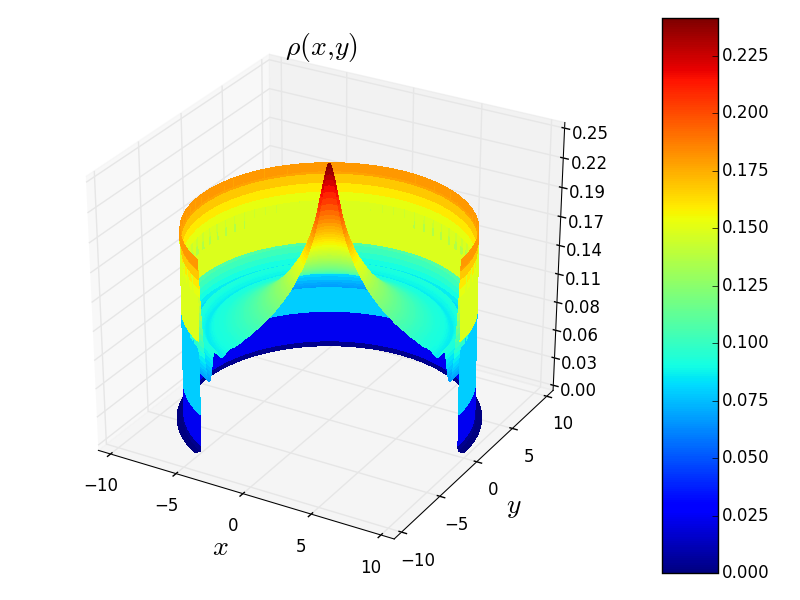}
\caption{
Three-dimensional visualization of the steady state data shown in 
Fig.~(\ref{rho_pipe_phi_0_05_Pe1250_Mono_hSph_f_t_sec2}). 
Peclet number $Pe=250$.
}
\label{rho_pipe_eq_phi_0_05_Pe1250_Mono_hSph_3D}
\end{figure}

In Fig.~(\ref{rho_pipe_eq_phi_0_05_Pe1250_Mono_hSph_3D}) we show a three-dimensional representation of 
the steady state profile (corresponding to the light-blue line in 
Fig.~(\ref{rho_pipe_phi_0_05_Pe1250_Mono_hSph_f_t_sec2})). 
Due to the cylindrical symmetry of the problem this representation does not provide additional information. 
However, in our opinion, it does provide a stronger intuitive impression of how the particles are 
distributed within the pipe. 
An important conclusion to be drawn from these simulations is that establishing a steady state 
migration peak requires approximately two orders of magnitude more time than the local rearrangements 
close to the wall. There is a clear separation of time scales. 
Although we have only shown data for a single volume fraction, we have checked that the same 
conclusion can be drawn over a wide range of volume fractions and shear-rates. 

\begin{figure}[!t]
\centering			
\includegraphics[width=0.5\textwidth]{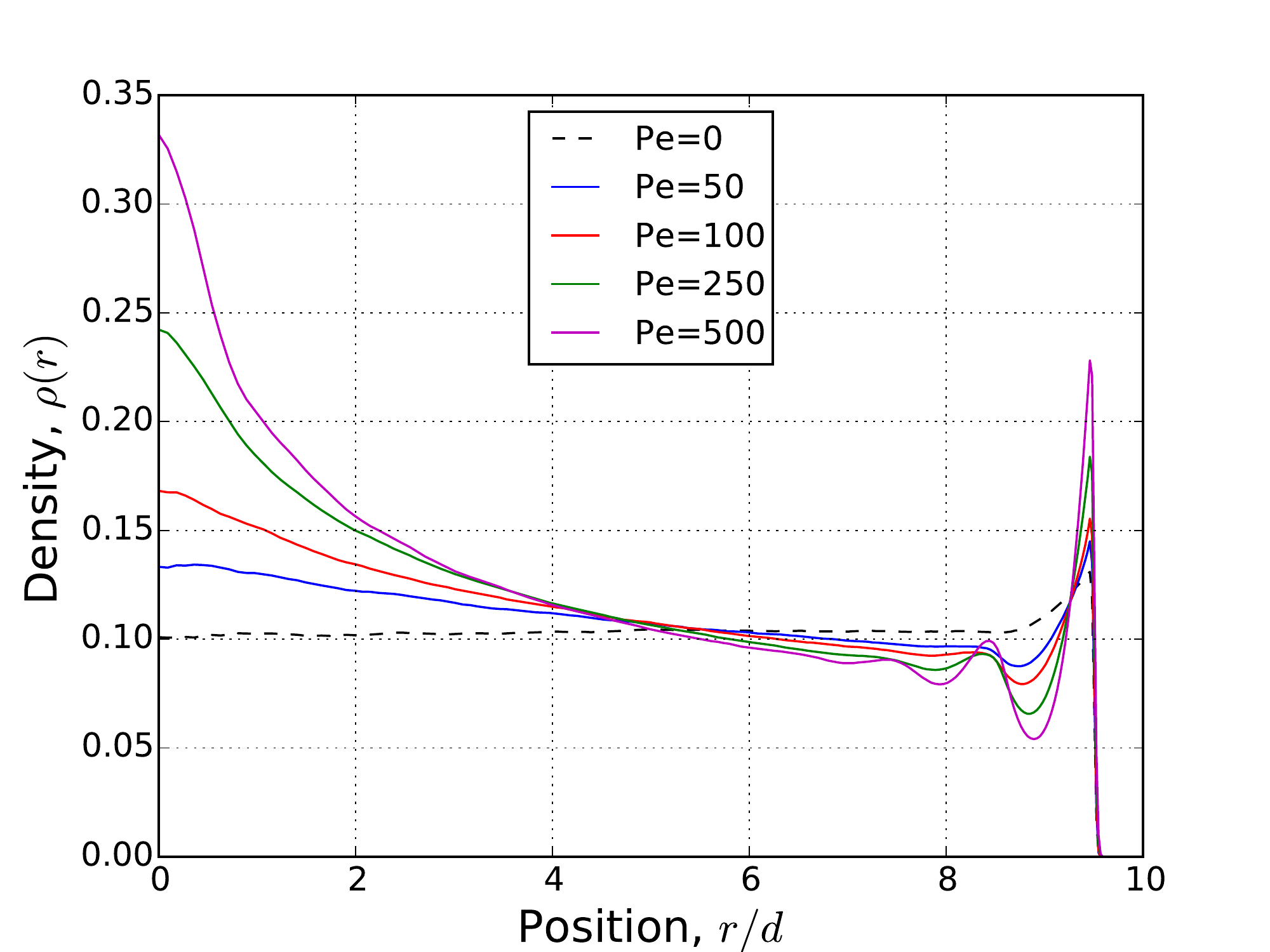}
\caption{The Peclet number dependence of the steady state density for 
monodisperse repulsive spheres confined in a cylindrical geometry.}
\label{rho_of_r_mono_diff_Pe}
\end{figure}

We next consider the influence of the shear-rate on the steady state density profiles. 
In Fig.~(\ref{rho_of_r_mono_diff_Pe}) we report the steady state density profiles for Peclet 
numbers ranging from $Pe=0$ to $500$. 
As expected, both the layering at the wall and the migration effect are enhanced by 
increasing the Peclet number, however, we also observe that the functional form of the profile 
changes in character in going from low to  high $Pe$. 
At low values of $Pe$ the decrease in density from the pipe centre towards the wall is an approximately 
linear function of the radial position, 
whereas for values of $Pe$ greater than around $200$ a significant curvature develops. 
For the volume fraction under consideration, $\phi=0.05$, we observe that the central peak height increases 
sublinearly as a function of $Pe$, eventually saturating at very high values of $Pe$ 
(although very large Peclet numbers were not systematically investigated).    
While these observations remain valid for low to intermediate volume fraction suspension, 
for dense liquid states ($\phi\ge0.4$) the picture becomes different. 
For such dense systems the density around the pipe centre becomes much less sensitive to 
increasing $Pe$, as many-particle packing effects interfere with and disrupt the migration 
mechanism~\cite{mig10}. 

\begin{figure}[!t]
\centering			
\includegraphics[width=0.5\textwidth]{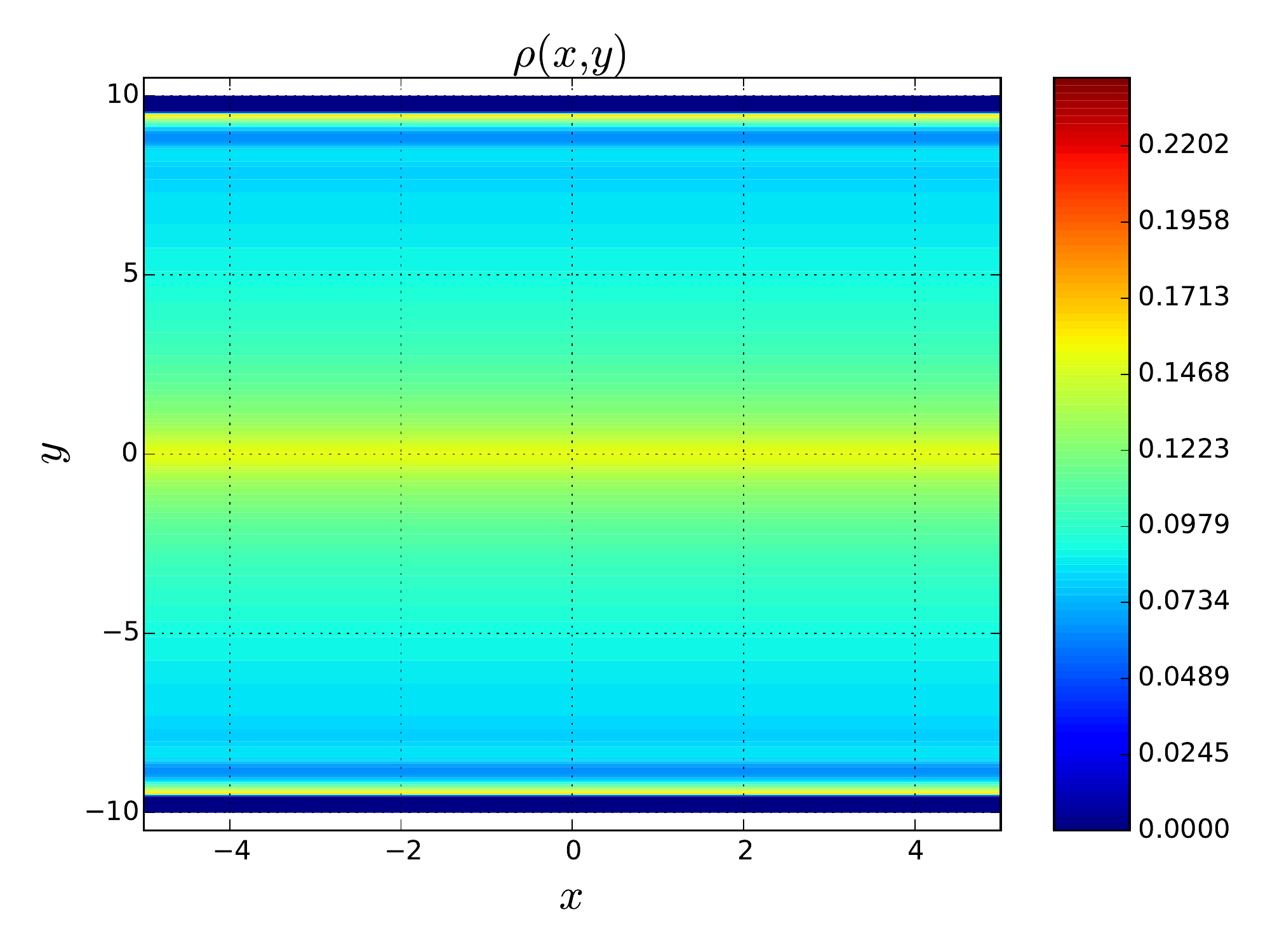}
\caption{A two dimensional representation (projection) of our three dimensional data for 
the steady state density profile in a channel geometry with flow in the $x$-direction 
and shear gradient in the $y$-direction. $Pe=250$.}
\label{channel_2D}
\end{figure}

\begin{figure}[!b]
\centering			
\includegraphics[width=0.5\textwidth]{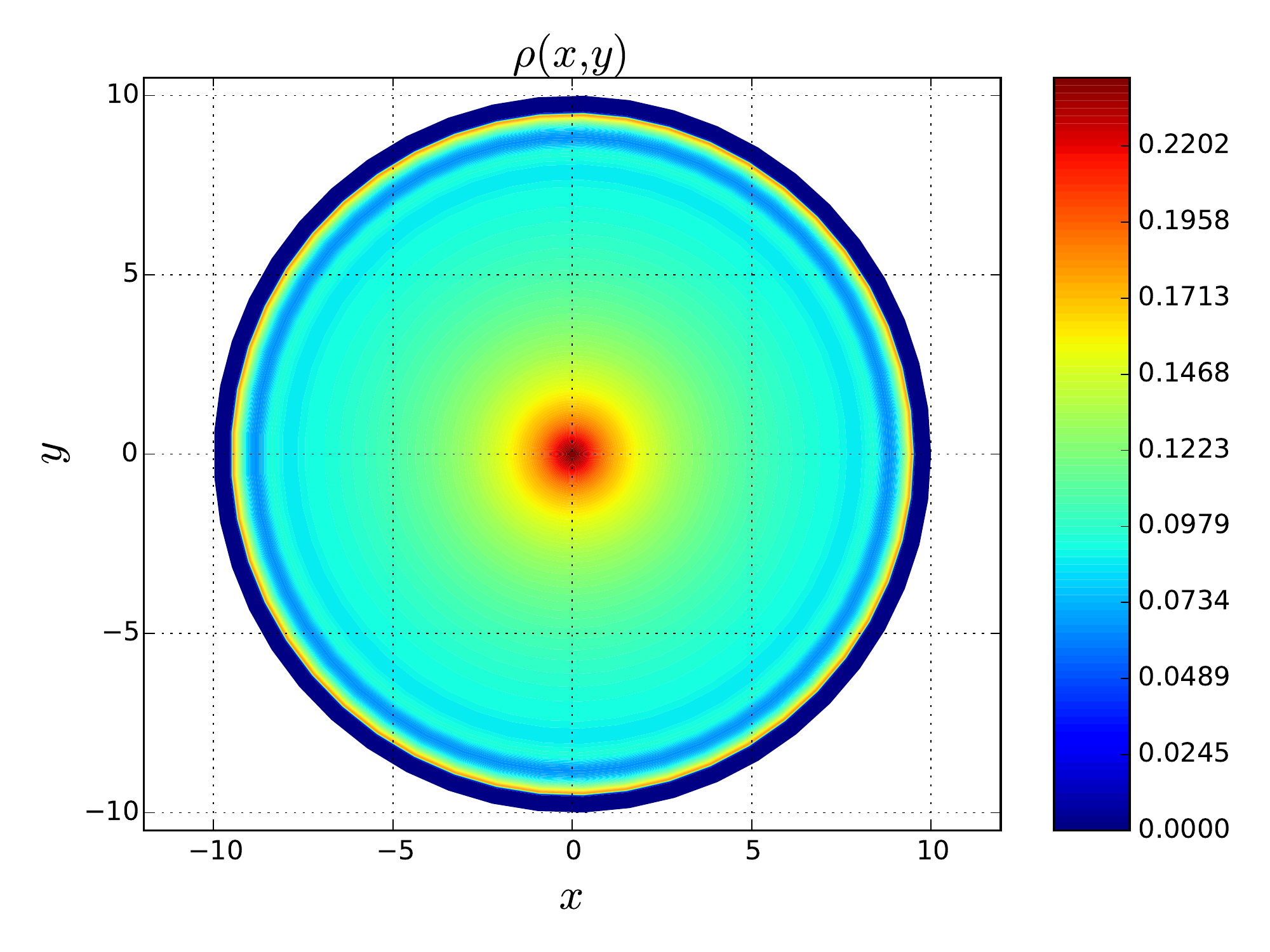}
\caption{
A two dimensional representation (projection) of our three dimensional data for 
the steady state density profile in a cylindrical geometry. $Pe=250$.
}
\label{pipe_2D}
\end{figure}

As the two most commonly studied, and physically relevant, geometries are cylindrical 
and planar (slit) confinement, it is of interest to compare the density profiles obtained 
in each case for equal values of the simulation parameters. 
In order to have a significant migration peak in steady-state 
we will fix the Peclet number for this comparison to the value $Pe=250$. 
The cylindrical system remains the same as that reported above. 
For the simulation of the planar system we have used the following parameters: the number of particles $N=250$, the box size $L_x=10d$, $L_y=20d$ and $L_z=13.09d$ and again a Brownian time step 
$dt=2\times 10^{-5}$. The external flow acts in the $z$-direction with shear-gradient in the 
$y$-direction. 
In Fig.~(\ref{channel_2D}) we show the projected steady state profile in the velocity-gradient 
plane. 
As expected, in the planar geometry the colloids are driven towards the central ($y=0$) 
line by the migration mechanism. 
As a phenomenological comparison, we report in Fig.~(\ref{pipe_2D}) the same situation for the system 
under calindrical confinement. 
It is apparent from the color bar, that the migration is relatively stronger in the 
cylindrical geometry than it is in the channel geometry. 
This effect is due to the fact that the migrational drift of the particles is `focused' on a 
central point (in our projected representation) instead of on a central line. 

\subsection{Bidisperse and polydisperse systems}\label{poly_disperse}

We now consider the case of a bidisperse system of spheres, interacting via the same pair potential 
as used in the previous section, but now with two species of different diameter, $d_1$ and $d_2$. 
The simulation parameters used here are: the total number of particles $N=400$, 
the Brownian time step $dt=5\times 10^{-6}$, the length of pipe $L=13.125d$, the maximal fluid 
velocity $v_m=2500$, number of particles of species 1 $N_1=200$ and the number of 
particle of species 2 $N_2=200$. 

\begin{figure}[!t]
\centering	
\includegraphics[width=0.5\textwidth]{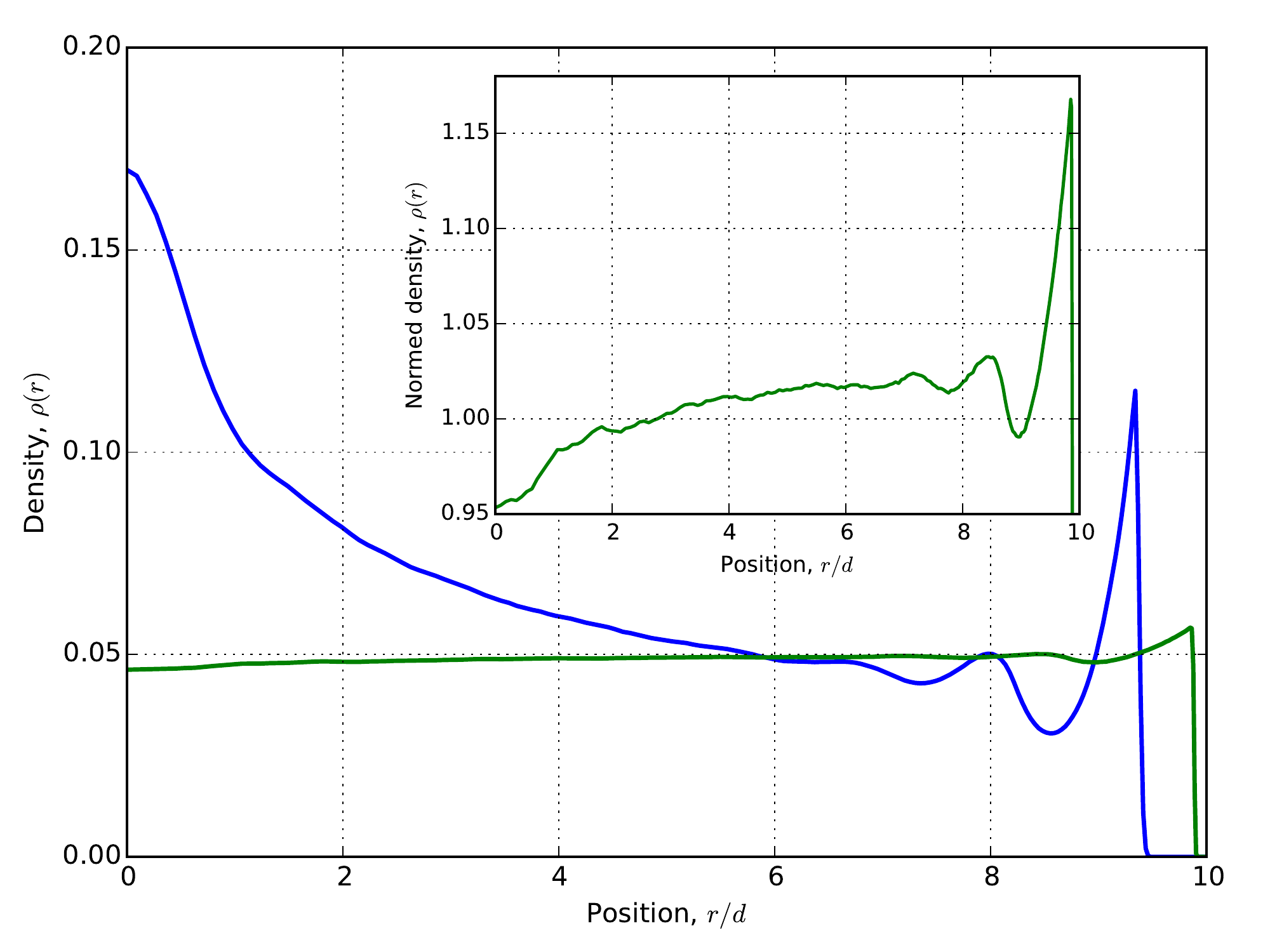}
\caption{In blue, the size class $d_2=1.25d$ and in green the size class $d_1=0.25d$. In the inset, the density profile of the class $d_1$, renormalised by the bulk density.}
\label{hard_spheres_bi}
\end{figure}

An interesting possibility which arises for this system is whether there exist conditions for which 
the migration effect can be reversed for one of the species, thus generating a steady-state with a 
density minimum at the center of the pipe. This effect has been demonstrated experimentally by Aarts {\it et al.}~\cite{aarts} for a mixture of red blood cells and
platelets.
Given that the driving force underlying the migration of particles is provided by an 
asymmetric distribution of colloidal collisions on the surface of a given particle, it follows 
intuitively that one would expect the larger of the two species to experience a stronger migration 
force than the smaller, which are consequently depleted from the region at the pipe center. 
These expectations are validated by the data shown in 
Fig.~(\ref{hard_spheres_bi}), for which we have simulated a system with 
$d_{1}=0.25d$ and $d_{2}=1.25d$. 
The main figure shows both density profiles on the same scale, 
whereas the inset focuses only on the second (smaller) species. 
The depletion in density at the center of the pipe is of the order of around $5\%$ of the bulk 
density for the parameters employed (the effect can be further enhanced by increasing the flow rate). 
We have found that decreasing the size-ratio of small to large spheres from unity to a value of 
around $0.2$ leads to an increase in the magnitude of the small particle depletion at the pipe 
centre. However, for smaller size ratios the large spheres are less efficient in excluding the 
smaller ones from the central region, as the latter can pass through the `gaps' between larger 
spheres, and the depletion effect becomes less prominent.     

From these results we can conclude that an anti-migration phenomenon can be 
obtained in bidisperse mixtures, provided that the size ratio, $d_2/d_1$, is not too small 
(smaller than around $0.2$).
This spatial separation of the two species could, for example, be exploited for the 
separation of biological fluids, where one could extract one species of particle at the 
boundary and the other from the center of the pipe.

We next consider a polydisperse system with a continuous distribution of particle sizes. 
For convenience we select the particle diameters from a Gaussian distribution with standard 
deviation $0.2d$ and average equal to unity. Due to the difference in the particle sizes 
one has to adapt the Brownian velocity accordingly. 
The standard deviation of the Brownian generator takes now the form 
$\hat{\sigma}_{i}=\sqrt{d_02T'dt/d_i}$, where $d_0$ is the reference diameter, here $d_0=1$, 
and $d_i$ is the diameter of particle $i$. 
The simulation parameters to be used for our polydisperse simulations are: number of 
particles $N=400$, length of the pipe $L=15.07d$, Brownian time step 
$dt=10^{-5}$ and the maximal velocity of the background fluid $v_m=2500$. 

\begin{figure}[!b]
\centering	
\includegraphics[width=0.5\textwidth]{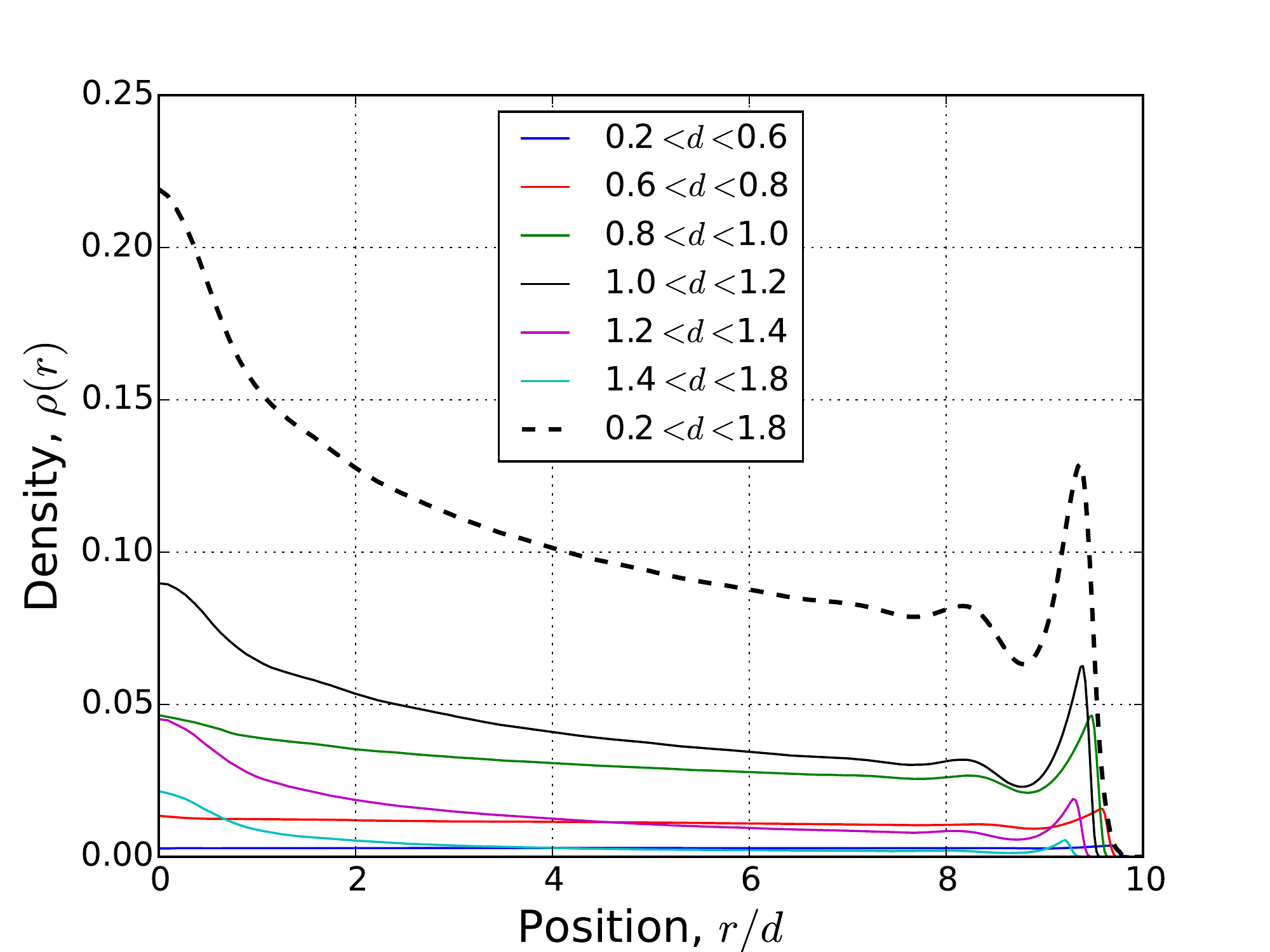}
\caption{Steady-state density profiles for polydisperse repulsive-spheres with 
polydispersity $0.2d$ and average diameter equal to unity under Poiseuille flow with 
maximal velocity $v_m=2500$.}
\label{rho_pipe_phi_0_05_Pe1250_gauss_steadyState_s02}
\end{figure}

As we are now dealing with a continuous size distribution it is useful to divide the particles 
into different size classes, e.g.~those particles with diameter between $0.8d$ and $d$, in order to 
aid the graphical presentation of our numerical data. 
In Fig.~(\ref{rho_pipe_phi_0_05_Pe1250_gauss_steadyState_s02}) we show the average density 
profiles for different size classes, together with the total density profile obtained by 
summing all of the individual classes. 
As expected, the total density displays a clear migration peak at the pipe centre, similar in 
form to the density of the monodisperse system at a comparable volume fraction. However, 
the oscillations close to the boundary, are less pronounced than in the monodisperse 
case, because the polydispersity frustrates the particle packing.

\begin{figure}[!t]
\centering	
\includegraphics[width=0.5\textwidth]{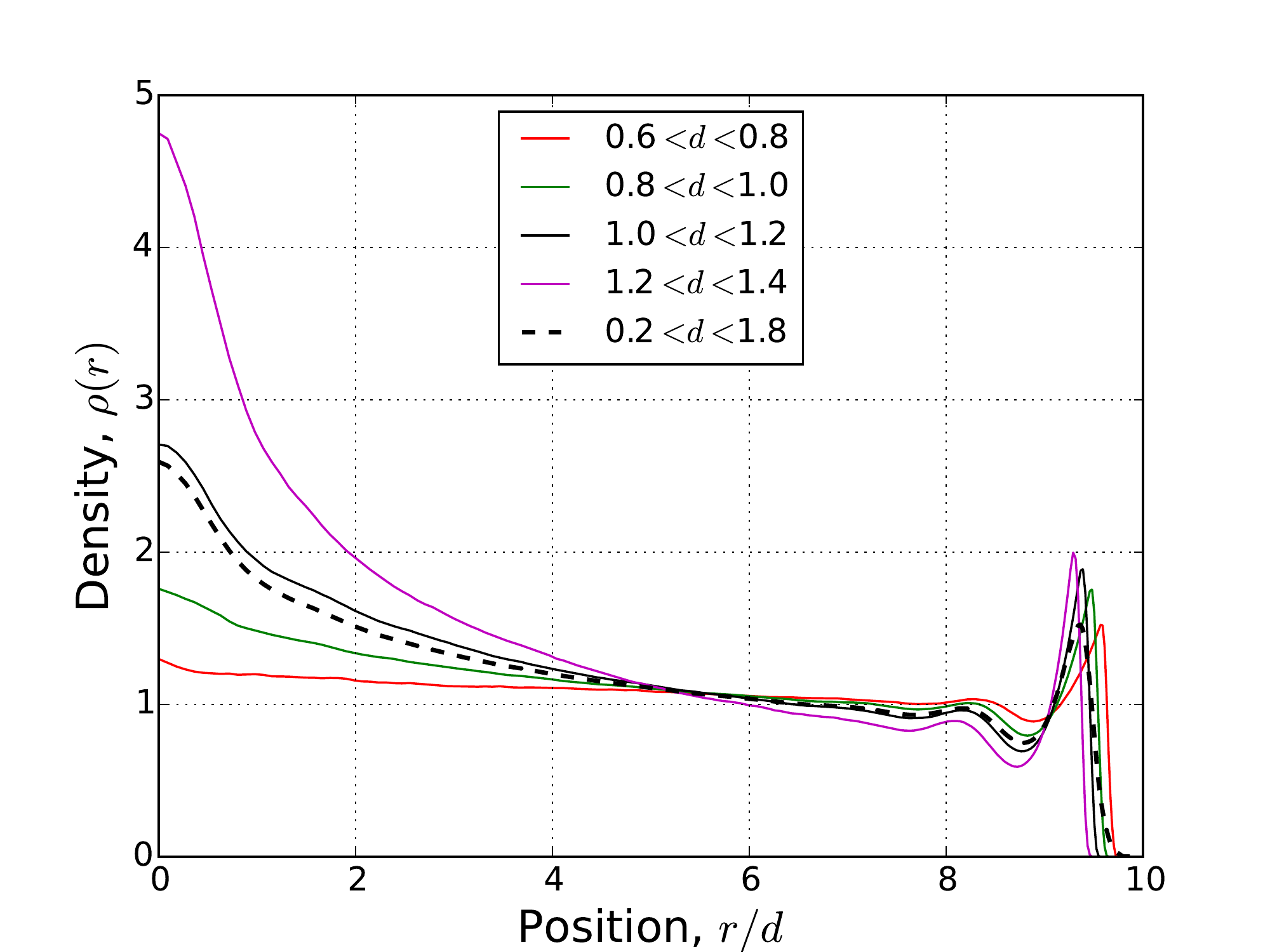}
\caption{The same data as shown in Fig.~(\ref{rho_pipe_phi_0_05_Pe1250_gauss_steadyState_s02}), 
normalized according to the bulk density of each size class. 
The central peak height decreases monotonically with decreasing particle size.}
\label{rho_pipe_phi_0_05_Pe1250_gauss_steadyState_normed_s02}
\end{figure}

The relative influence of migration on the different particle size classes is made more clear 
in Fig.~(\ref{rho_pipe_phi_0_05_Pe1250_gauss_steadyState_normed_s02}), where we show each of 
the profiles normalized by the appropriate bulk density. 
If we take the data for the total density (broken line in the figure) to represent the average 
migration profile of the particles, then it is clear that the classes 
containing the larger particles exhibit a much stronger migration peak than those containing 
the smaller particles. This is consistent with our findings from the bidosperse system. 
If we recall the definition of the Peclet number ($Pe=\frac{v_{m}d^2}{R D_{0}}$) we can 
rationalize these observations: the bigger particles experience a larger Peclet number than 
the smaller ones, and are thus driven more effectively to the center of the pipe.

\subsection{Mono-disperse Lennard-Jones spheres}\label{mono_disperse_LJ}

The purely repulsive interactions so far considered can be realized using certain, specially prepared 
model systems (e.g.~PMMA spheres). However, in general, colloids in suspension also exhibit an 
attractive component to their interaction potential. 
It is thus somewhat surprising that to date there have been no studies of the influence of 
interparticle attraction on the steady state density profiles in Poiseuille flow. 
In equilibrium, attractive systems demonstrate phase separation into colloid-rich and colloid-poor 
phases. In the present nonequilibrium situation it is not clear how the cohesive tendency of the 
particles will interact with the migration mechanism. On one hand, the development of a migration 
peak might be expected to be favoured, as the formation of dense regions would lead to a lowering 
of the systems potential energy. On the other hand, it is not clear how the migration mechanism itself 
(which generates the density peak) is altered when the particles prefer to stick together: an 
excess of collisions on one side of a particle surface could give rise to a `pull' towards the walls, 
rather than an inwards `push', as is the case for purely repulsive interactions.

\begin{figure}[!b]
\centering	
\includegraphics[width=0.5\textwidth]{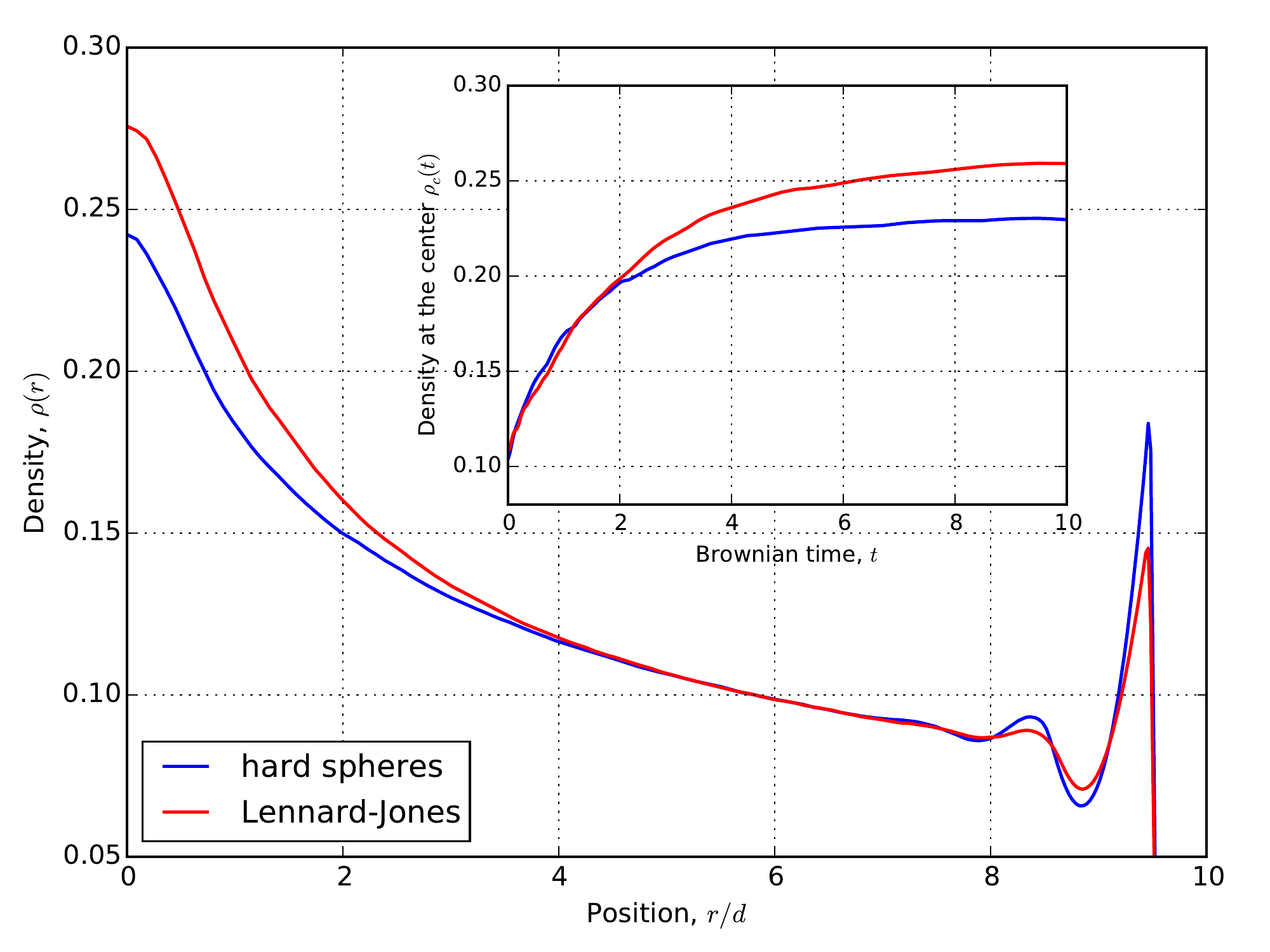}
\caption{Comparison of the steady-state density profile of repulsive particles with that 
of attractive Lennard-Jones particles. 
The attraction parameter $\epsilon=0.5$ and the Poiseuille flow is the 
same as that considered in the previous figures. 
In the inset we show the height of the central peak as a function of the time 
since the onset of flow.
}
\label{lennard_jones_mono}
\end{figure}

The simulations have been performed using the following parameters: 
the number of particles $N=400$, 
the length of the pipe $L=13.33d$ and the Brownian time step $dt=2\times 10^{-5}$. 
The interparticle interaction potential takes the well-known form
\begin{equation}
U_{LJ}=4\epsilon\left[\left(\frac{d}{r}\right)^{12}-\left(\frac{d}{r}\right)^{6}\right].
\end{equation}
For the wall-particle interaction we retain the same repulsive potential used previously. 
In Fig.~(\ref{lennard_jones_mono}) we show the steady-state density profile of a system formed by 
pseudo-hard spheres and Lennard-Jones spheres. 
The main observation to be made here is that the central migration peak is higher for attractive 
particles than for repulsive particles. 
This enhancement can be made larger by increasing $\varepsilon$ (the attraction strength) beyond the 
value $\varepsilon=0.5$ considered in Fig.~(\ref{lennard_jones_mono}). However, one should be careful 
to remain at $\varepsilon$ values away from the two-phase region of the Lennard-Jones phase diagram.  
Increasing the value beyond $\epsilon\approx 1$ complicates the situation by incorporating the 
physics of phase separation dynamics. Although we have not studied these effects in detail we do 
find significant sensitivity of the steady state profiles to $\epsilon$ in the vicinity of 
the binodal. 

In the inset of Fig.~(\ref{lennard_jones_mono}) we show the value of the central density peak 
as a function of the time since the onset of flow. 
The steady state 
is achieved slightly faster for the repulsive system than for the attractive system. 
We note that the initial growth of the peak height is virtually identical for the 
first two Brownian time units. 
Recalling that we are working at rather low volume fractions, 
this suggests that the mechanism of migration, 
on the level of binary collisions, is not significantly different in the two systems. 
However, for later times the density in the central region becomes sufficiently large that 
the interparticle attraction causes the particles to pull together, enhancing the height of 
the peak. 

\begin{figure}[!t]
\centering	
\includegraphics[width=0.5\textwidth]{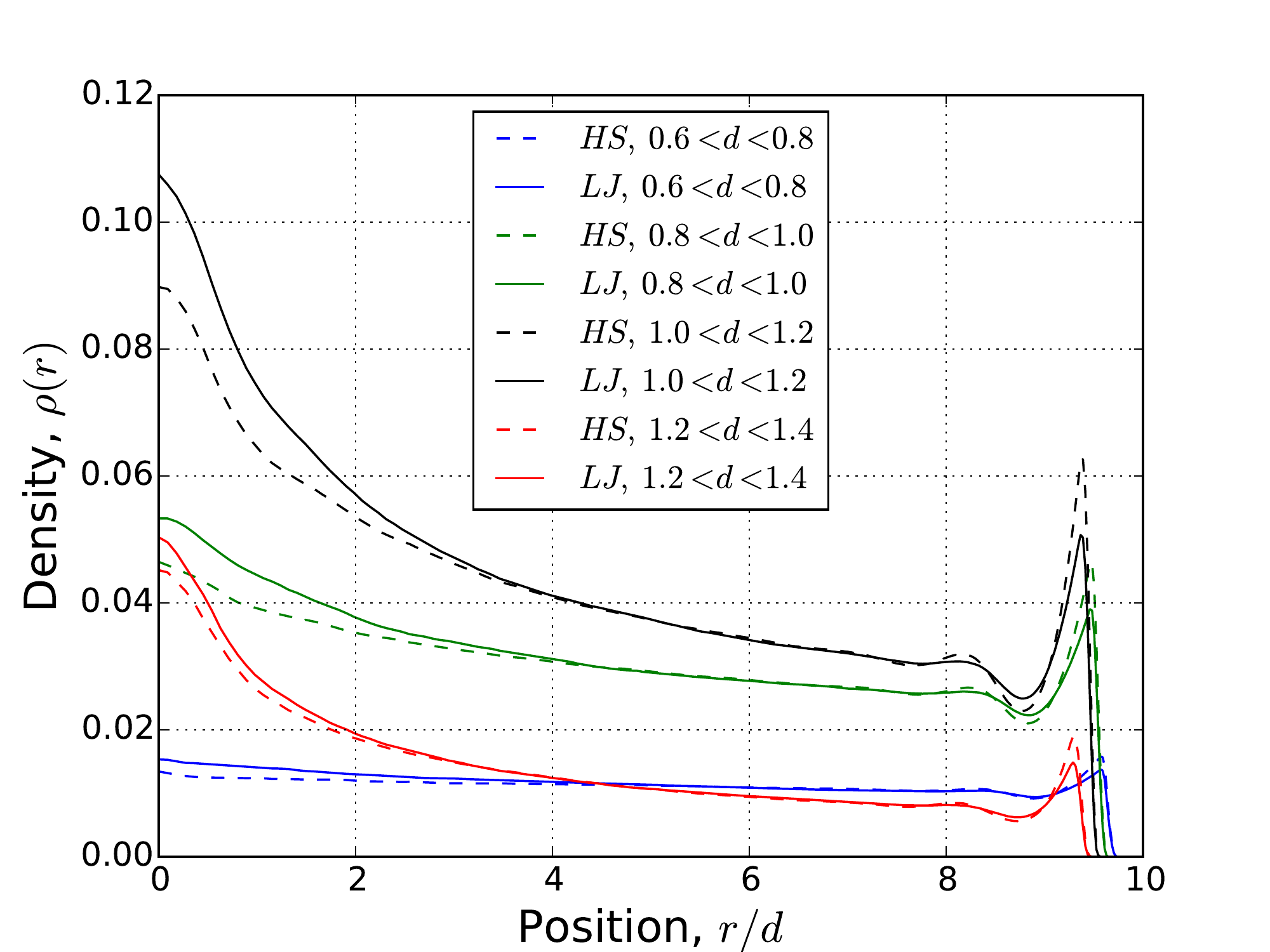}
\caption{Comparison of the steady state density distribution of a system of 
polydisperse, attractive (Lennard-Jones) particles with that of a system of 
polydisperse, repulsive particles. Both systems are characterised by a 
polydispersity $0.2d$ and average of $d$. 
The Lennard-Jones attraction parameter is set to $\epsilon=0.5$.}
\label{lennard_jones_poly}
\end{figure}

\subsection{Polydisperse Lennard-Jones spheres}\label{poly_disperse_LJ}
In order to bring our investigations somewhat closer to experimental reality, we now consider 
the steady-state profiles of polydisperse Lennard-Jones (LJ) particles. 
As previously, we will employ a Gaussian distribution of particle sizes and compare 
our results for the attractive system with those previously obtained for the purely repulsive 
system. 
The results are represented in Fig.~(\ref{lennard_jones_poly}) for a maximal velocity 
of $v_m=2500$. 

For each size class the migration peak is enhanced by turning on the attraction, but this enhancement 
is most pronounced for the larger particles. 
This suggests that interparticle attraction could be used to increase yield efficiency 
of devices which segregate the particles according to their size.
The first packing peak at the wall decreases with increasing attraction, because our choice of 
repulsive substrate tends towards drying, rather than wetting, as the attraction is increased.



\section{Conclusions}\label{conclusion}

In this paper we have presented a simulation study of shear induced migration in 
monodisperse, bidisperse and polydisperse systems subject to Poiseuille flow. 
For repulsive interparticle interactions we find that the migration induced density peak 
at the center of the channel develops much more slowly (around two orders of magnitude, for 
the parameters considered) than local structural changes close to the channel boundaries. 
For monodisperse systems we have shown that the steady state accumulation of particles 
at the center of the channel is more pronounced in cylindrical geometry than in planar 
geometry. 
For systems of polydisperse repulsive particles we find that the migration effect is 
strongest for the larger particles, suggesting that migration could possibly be expoited 
as a mechanism for sorting or separating particles according to their size. 
In the special case of a bidisperse system we find that it is possible to induce an 
`anti-migration' effect in the smaller size particles, whereby their density is depleted, 
rather than enhanced, in the central region of the channel. 

By considering systems of Lennard-Jones particles we have addressed the influence of 
interparticle attractions on the steady state density profiles. We note that care was 
taken to consider only parameters for which bulk phase separation does not occur. 
We find that adding an interparticle attraction enhances the height of the central density 
peak in steady state, both for monodisperse and polydisperse systems, and could then be used 
as a means to optimise particle segregation effects in potential applications. 
However, the transient dynamics in going from equilibrium to steady state are not 
strongly influenced by the presence or absence of attraction. 

There are several possibilities to extend the findings of the present work. 
In Refs.~\cite{brader_kruger,scacchi_krueger_brader2016} a dynamical density functional 
theory was developed, which incorporates the physics of shear-induced migration in a 
mean-field fashion. It would be interesting to exploit the numerical data presented in this 
work to optimize the approximations employed in the theory - particularly for the case 
of attractive interparticle interactions. 
In this work we have focused exclusively on systems with a strongly repulsive core. However, 
there is also much interest in model systems for which the particles interact via soft, 
penetrable interactions (modelling, e.g.~polymer coils). Whether the picture of migration 
presented here holds also for these systems remains an open question. Finally, it seems to us 
possible that, should the migration lead to a sufficiently enhanced density at the channel 
center, then crystallization effects could start to play a role: Can a steady state crystal 
structure be induced by non uniform flow? We will continue to investigate these and related  
phenomena.

\section{Acknowledgement}
We thank the Swiss National Science Foundation for financial support. 
\\


\begin{thebibliography}{0}
\expandafter\ifx\csname natexlab\endcsname\relax\def\natexlab#1{#1}\fi
\expandafter\ifx\csname bibnamefont\endcsname\relax
  \def\bibnamefont#1{#1}\fi
\expandafter\ifx\csname bibfnamefont\endcsname\relax
  \def\bibfnamefont#1{#1}\fi
\expandafter\ifx\csname citenamefont\endcsname\relax
  \def\citenamefont#1{#1}\fi
\expandafter\ifx\csname url\endcsname\relax
  \def\url#1{\texttt{#1}}\fi
\expandafter\ifx\csname urlprefix\endcsname\relax\def\urlprefix{URL }\fi
\providecommand{\bibinfo}[2]{#2}
\providecommand{\eprint}[2][]{\url{#2}}

\end{thebibliography}


\begin{thebibliography}{}



\bibitem{brader_review}
J.M.~Brader, J.Phys.:Condens.Matter {\bf 22} 363101 (2010). 

\bibitem{banding1}
J.K.G.~Dhont {\it et al.}, 
Faraday Discuss. {\bf 123} 157 (2003).

\bibitem{banding2}
R.~Besseling, L.~Isa, P.~Ballesta and W.~Poon, 
Phys. Rev. Lett. {\bf 105} 268301 (2010).

\bibitem{banding3}
H.~Jin  {\it et al.} 
Soft Matter {\bf 10} 9470 (2014).

\bibitem{banding4}
F.~Varnik {\it et al.} 
Phys. Rev. Lett. {\bf 90} 095702 (2003).

\bibitem{blood}
D.~Katanov, G.~Gompper and d.A.~Fedosov, 
Microvascular Research {\bf 99} 57 (2015).

\bibitem{brader_kruger}
J.M.~Brader and M.~Kr\"uger, 
Mol.Phys. {\bf 109} 1029 (2011). 

\bibitem{lane_brady}
D.R.~Foss and J.R.~Brady, 
J. Rheol. {\bf 44} 629 (2000).

\bibitem{mig1}
D.Semwogerere and E.R.Weeks, Phys.Fluids {\bf 20} 043306 (2008).

\bibitem{mig2}
A.Kumar and M.D.Graham, Soft Matter {\bf 8} 10536 (2012).

\bibitem{mig3}
K.L.McCarthy and W.L.Kerr, J.Food Eng. {\bf 37} 11 (1998).

\bibitem{gadala}
F.~Gadala-Maria and A.~Acrivos, 
J. Rheol. {\bf 24}, 799 (1980). 

\bibitem{mig4}
D.Leighton and A.Acrivos, J.Fluid.Mech. {\bf 181} 415 (1987).

\bibitem{nozieres}
P.~Nozieres, D.~Quemada, Europhys. Lett. {\bf 2} 129 (1986). 

\bibitem{mig5}
R.J Phillips et al., Phys.Fluids A {\bf 4} 30 (1992).

\bibitem{scacchi_krueger_brader2016}
A.~Scacchi, M.~Kr\"uger and J.M.~Brader, 
J.Phys.: Condens. Matter {\bf 28} 244023 (2016).



\bibitem{mig6}
P.R.Nott and J.F.Brady, J.Fluid Mech. {\bf 275} 157 (1994).

\bibitem{mig7}
J.F.Morris and J.F.Brady, Int.J.Multiph.Flow {\bf 24} 105 (1998).

\bibitem{mig8}
A.Acrivos, R.Mauro and X.Fan, Int.J.Multiphase Flow {\bf 19} 797 (1993).

\bibitem{mig9}
M.Frank, D.Anderson, E.R.Weeks and J.F.Morris, J.Fluid.Mech. {\bf 493} 363 (2003).

\bibitem{mig10}
M.K.Lyon and L.G.Leal, J.Fluid.Mech. {\bf 363} 25 (1998).


\bibitem{mig11}
M.K.Lyon and L.G.Leal, J.Fluid.Mech. {\bf 363} 57 (1998).

\bibitem{allen}
M.P.~Allen and D.J.~Tildesley, 
{\it Computer simulation of liquids} (Clarenden Press, Oxford, 1989). 

\bibitem{aarts}
P.A.M.M.~Aart, S.A.T.~van den Broek, G.W.~Prins, G.D.C.~Kuiken, J.J.~Sixma, R.M.~Heethaar, Arteriosclerosis, Thrombosis, and Vascular Biology, {\bf 8} 819 (1988)



\end{thebibliography}
\end{document}